\PassOptionsToPackage{colorlinks=true,linkcolor=blue,filecolor=black,urlcolor=blue,citecolor=magenta}{hyperref}
\documentclass[11pt]{article}
\usepackage{myjheppub}
\usepackage{amsmath,amssymb}
\usepackage{mathrsfs}
\usepackage{amsfonts}
\usepackage{amsthm}
\usepackage[table]{xcolor}
\usepackage{caption,subcaption}
\usepackage{tikz}
\usetikzlibrary{decorations.markings}
\usetikzlibrary{patterns}
\usepackage{color, soul}
\usepackage{feynmp-auto}
\usepackage{slashed}
\usepackage{arydshln}

\DeclareMathOperator{\arccosh}{arccosh}

\numberwithin{equation}{section}

\title{ 
	 More on the Subleading Gravitational Waveform
	}
\author[a]{Alessandro Georgoudis}
\author[b]{Carlo Heissenberg}
\author[c]{Ingrid Vazquez-Holm}

\affiliation[a]{
	Centre for Theoretical Physics, Department of Physics and Astronomy,
	Queen Mary University of London, Mile End Road, London, E1 4NS, United Kingdom}
\affiliation[b]{
	School of Mathematical Sciences,
	Queen Mary University of London, Mile End Road,	London, E1 4NS, United Kingdom}
\affiliation[c]{Department of Physics and Astronomy, Uppsala University, Box 516, 75120 Uppsala, Sweden
}

\abstract{
In this short note, we address the calculation of the contribution to the one-loop gravitational waveform 
arising from the difference of the unitarity cuts associated to the $s$- and $s'$-channels
recently pointed out in 
Ref.~\cite{Caron-Huot:2023vxl}, 
providing its explicit expression for minimally-coupled massive scalars in momentum space.
}

\begin{document}

\maketitle
\section{Amplitude and Waveform at Subleading Order}
\label{sec:intro}

The quest for improving the understanding of gravitational-wave emissions from unbound binaries in the post-Minkowskian expansion using scattering amplitudes \cite{Bjerrum-Bohr:2018xdl,Cheung:2018wkq,Kosower:2018adc,Bern:2019nnu,KoemansCollado:2019ggb,Bern:2019crd,Parra-Martinez:2020dzs,DiVecchia:2020ymx,DiVecchia:2021ndb,Bern:2021dqo,Herrmann:2021lqe,DiVecchia:2021bdo,Herrmann:2021tct,Heissenberg:2021tzo,Bjerrum-Bohr:2021vuf,Bjerrum-Bohr:2021din,Cristofoli:2021vyo,Damgaard:2021ipf,Brandhuber:2021eyq,Bjerrum-Bohr:2021wwt,Bern:2021yeh,Cristofoli:2021jas,Manohar:2022dea,DiVecchia:2022owy,DiVecchia:2022nna,DiVecchia:2022piu,DiVecchia:2023frv,Damgaard:2023vnx,Damgaard:2023ttc} and worldline methods inspired by quantum field theory \cite{Goldberger:2016iau,Kalin:2020mvi,Kalin:2020fhe,Mogull:2020sak,Jakobsen:2021smu,Mougiakakos:2021ckm,Dlapa:2021npj,Riva:2021vnj,Dlapa:2021vgp,Jakobsen:2022psy,Kalin:2022hph,Dlapa:2022lmu,Dlapa:2023hsl} has witnessed recent progress in the calculation of the $2\to3$ amplitude for the scattering of two massive scalars and a single graviton emission up to one loop \cite{Brandhuber:2023hhy,Herderschee:2023fxh,Elkhidir:2023dco,Georgoudis:2023lgf}, schematically 
$\mathcal A =\mathcal A_0+\mathcal A_1 +\mathcal O(G^{7/2})$ in terms of Newton's constant $G$.
In  Ref.~\cite{Georgoudis:2023lgf}, we have shown that, in accordance with unitarity, the one-loop amplitude takes the following form in the classical, or near-forward, limit $\hbar\to0$, where $\hbar$ is used as a bookkeeping device to keep track of the scaling of the momentum transfer,
\begin{equation}\label{amplitude}
	\mathcal A_1 = \mathcal B_1+\frac{i}{2}\left(s+s'\right)+\frac{i}{2}\left(c_1+c_2\right)\,.
\end{equation}
Here, $\mathcal B_1$ denotes the real part of $\mathcal A_1$ while $s$, $s'$, $c_1$, $c_2$ are the unitarity cuts defined by
\begin{equation}\label{Schannel}
	s=
	\begin{gathered}
		\begin{tikzpicture}
			\path [draw, ultra thick, blue] (-4,2)--(-.3,2);
			\path [draw, ultra thick, green!60!black] (-4,1)--(-.3,1);
			\path [draw, red] (-1,1.5)--(-.32,1.5);
			\filldraw[black!20!white, thick] (-3,1.5) ellipse (.5 and .8);
			\draw[thick] (-3,1.5) ellipse (.5 and .8);
			\filldraw[black!20!white, thick] (-1.3,1.5) ellipse (.5 and .8);
			\draw[thick] (-1.3,1.5) ellipse (.5 and .8);
		\end{tikzpicture}
	\end{gathered}
	\qquad\quad
	s'=
	\begin{gathered}
		\begin{tikzpicture}
			\path [draw, ultra thick, blue] (-4,2)--(-.3,2);
			\path [draw, ultra thick, green!60!black] (-4,1)--(-.3,1);
			\path [draw, red] (-3,1.5)--(-2.1,1.5);
			\filldraw[black!20!white, thick] (-3,1.5) ellipse (.5 and .8);
			\draw[thick] (-3,1.5) ellipse (.5 and .8);
			\filldraw[black!20!white, thick] (-1.3,1.5) ellipse (.5 and .8);
			\draw[thick] (-1.3,1.5) ellipse (.5 and .8);
		\end{tikzpicture}
	\end{gathered}
\end{equation}
and
\begin{equation}\label{Cchannel}
	c_1=
	\begin{gathered}
		\begin{tikzpicture}
			\path [draw, ultra thick, blue] (-4,2)--(-.3,2);
			\path [draw, ultra thick, green!60!black] (-4,1)--(-2.1,1);
			\path [draw, red] (-3,1.5)--(-.3,1.5);
			\filldraw[black!20!white, thick] (-3,1.5) ellipse (.5 and .8);
			\draw[thick] (-3,1.5) ellipse (.5 and .8);
			\filldraw[black!20!white, thick] (-1.3,1.75) ellipse (.45 and .55);
			\draw[thick] (-1.3,1.75) ellipse (.45 and .55);
		\end{tikzpicture}
	\end{gathered}
	\qquad\quad
	c_2=
	\begin{gathered}
		\begin{tikzpicture}
			\path [draw, ultra thick, blue] (-4,2)--(-2.1,2);
			\path [draw, ultra thick, green!60!black] (-4,1)--(-.3,1);
			\path [draw, red] (-3,1.5)--(-.3,1.5);
			\filldraw[black!20!white, thick] (-3,1.5) ellipse (.5 and .8);
			\draw[thick] (-3,1.5) ellipse (.5 and .8);
			\filldraw[black!20!white, thick] (-1.3,1.25) ellipse (.45 and .55);
			\draw[thick] (-1.3,1.25) ellipse (.45 and .55);
		\end{tikzpicture}
	\end{gathered}
\end{equation}
While $\mathcal B_1$, $c_1$ and $c_2$ scale classically, $\mathcal O(\hbar^{-1})$, we found that $s+s'$ only contributes to superclassical order, $\mathcal O(\hbar^{-2})$ and does not give any classical $\mathcal O(\hbar^{-1})$ contribution. There is perfect agreement between the results of Refs.~\cite{Brandhuber:2023hhy,Herderschee:2023fxh,Georgoudis:2023lgf} for \eqref{amplitude} evaluated on numerical points.

Following the Kosower--Maybee--O'Connell (KMOC) strategy \cite{Kosower:2018adc}, one can show that the waveform kernel $W = W_0+W_1+\cdots$, whose Fourier transform is the frequency-domain waveform, is given to leading order \cite{Cristofoli:2021vyo} by $W_0 = \mathcal A_0$  and to subleading order \cite{Caron-Huot:2023vxl} by
\begin{equation}\label{kernel}
	W_1 = \mathcal B_1+\frac{i}{2}\left(s-s'\right)+\frac{i}{2}\left(c_1+c_2\right)\,.
\end{equation}
The relative minus sign between the $s$ and $s'$ in the one-loop kernel \eqref{kernel} is crucial to ensure that their superclassical $\mathcal O(\hbar^{-2})$ terms cancel out against each other, while they add up in the amplitude \eqref{amplitude}, as predicted by the eikonal exponentiation \cite{Damgaard:2021ipf,Cristofoli:2021jas,DiVecchia:2022piu} and checked in \cite{Georgoudis:2023lgf}. However, $s-s'$ had been assumed to vanish at $\mathcal O(\hbar^{-1})$ as well, in analogy with $s+s'$ which appears in the amplitude, and was consequently dropped from the waveform kernel \eqref{kernel}. This is however not correct, and the need to also include it to obtain the complete KMOC kernel at classical $\mathcal O(\hbar^{-1})$ order was pointed out in Ref.~\cite{Caron-Huot:2023vxl}. In the post-Newtonian \cite{Bini:2023fiz} and soft \cite{Aoude:2023dui} limits, it was noted that the omission of this term can be compensated for by a suitable redefinition of the reference frame, a mechanism further generalized in \cite{Georgoudis:2023eke} to generic velocities and frequencies. The purpose of this note is to present the explicit calculation of this extra piece in momentum space.

\section{Unitarity Cuts Revisited}
\label{sec:waveform}

Following Refs.~\cite{Kosower:2018adc,Cristofoli:2021vyo,Cristofoli:2021jas}, we consider the expectation value of the graviton field,
\begin{equation}\label{}
	H_{\mu\nu}(x) = \int_k \left[
	e^{ik\cdot x} a_{\mu\nu}(k)
	+
	e^{-ik\cdot x} a^{\dagger}_{\mu\nu}(k)
	\right]
	,\qquad
	\int_k \equiv
	\int 2\pi\delta(k^2) \theta(k^0)
	\frac{d^Dk}{(2\pi)^D}\,,
\end{equation}
in the state $|\text{out}\rangle  = S |\text{in} \rangle$ obtained by applying the $S$ matrix to an in-state modeling two incoming objects with masses $m_1$, $m_2$, velocities $u_1^\mu$, $u_2^\mu$, so that $y=-u_1\cdot u_2$, and impact parameters $b_1^\mu$, $b_2^\mu$, so that $b^\mu=b_1^\mu-b_2^\mu$ and $b\cdot u_{1,2}=0$, in terms of wavepackets.
This expectation is related to the metric fluctuation by
\begin{equation}\label{}
	\frac{g_{\mu\nu}(x)-\eta_{\mu\nu}}{\sqrt{32\pi G}}
	=
	h_{\mu\nu}(x) = \langle \text{out} | H_{\mu\nu}(x) |\text{out}\rangle
	=
	\int_k
	e^{ik\cdot x}  \langle\text{out}|a_{\mu\nu}(k)|\text{out}\rangle
	+\text{c.c.}\,,
\end{equation}
where ``c.c'' means ``complex conjugate''.
Defining the Fourier transform
\begin{equation}\label{FT5}
	\begin{split}
		\operatorname{FT}\left[
		f^{\mu\nu}
		\right]
		=
		\tilde{f}^{\mu\nu}
		&=
		\int \frac{d^Dq_1}{(2\pi)^D}\frac{d^Dq_2}{(2\pi)^D}\,(2\pi)^D\delta^{(D)}(q_1+q_2+k) \\
		&\times
		2\pi\delta(2 m_1 u_1\cdot q_1)2\pi\delta(2 m_2 u_2\cdot q_2)
		e^{ib_1\cdot q_1+ib_2\cdot q_2} f^{\mu\nu}\,,
	\end{split}
\end{equation}
one finds
\begin{equation}\label{hmunuWmunu}
	h_{\mu\nu}(x)
	=
	\int_k\left[
	e^{ik\cdot x}\,
	i
	\operatorname{FT}
	\langle\, W_{\mu\nu} \rangle
	\right]
	+
	\text{c.c.}\,,\qquad
	W^{\mu\nu} = W_0^{\mu\nu} + W_1^{\mu\nu} +\mathcal O(G^{7/2})
\end{equation}
where $\langle\, \cdots \rangle$ is a shorthand notation for the wavepacket average, and $W_0=\mathcal A_0$ while $W_1$ is given by Eq.~\eqref{kernel}. Indeed, at this order, using $S = e^{iN} = 1+iN-\tfrac12 N^2+\cdots$,
\begin{equation}\label{aaNaNNNaN}
	\langle\text{out}|a_{\mu\nu}(k)|\text{out}\rangle
	=
	i \langle\text{in}|a_{\mu\nu}(k) N|\text{in}\rangle
	- \frac{1}{2} \langle\text{in}|a_{\mu\nu}(k) N^2|\text{in}\rangle
	+
	\langle\text{in}|N a_{\mu\nu}(k) N|\text{in}\rangle + \cdots\,.
\end{equation}
At one loop, the $N$ insertion simply leads to the real part of the kernel, $\mathcal B_1$ \cite{Georgoudis:2023lgf}, while, inserting a complete set of states, we see that $\langle\text{in}|a_{\mu\nu}(k) N^2|\text{in}\rangle$ includes the sum of all cuts, and $\langle\text{out}|N a_{\mu\nu}(k) N|\text{out}\rangle$ only contains $s'$. Thanks to the factor of $-\tfrac12$ in \eqref{aaNaNNNaN}, this leads to \eqref{kernel}.
The waveform kernel thus involves three building blocks: $\mathcal B_1$ and $\frac{i}{2}(c_1+c_2)$, which where calculated in Refs.~\cite{Brandhuber:2023hhy,Herderschee:2023fxh,Georgoudis:2023lgf}, and $\frac{i}{2}(s-s')$, which is instead new. 

We provide the expression of $\tfrac{i}{2}(s-s')$ at $\mathcal O(\hbar^{-1})$ in the ancillary files of this note, focusing on the case of minimally coupled massive scalars in general relativity, and including both the infrared (IR) divergent $\mathcal O(\epsilon^{-1})$ piece and the finite $\mathcal O(\epsilon^0)$ piece. To obtain this result, we explicitly  constructed the integrands of $s$ and $s'$ by suitably gluing the tree-level $2\to2$ and $2\to3$ amplitudes given e.g.~in Appendix~B of \cite{Georgoudis:2023lgf}, and performed the resulting integrals via reverse unitarity following \cite{Georgoudis:2023lgf}. 
We have also checked that the same result can be obtained by starting from the complete quantum integrand \cite{Carrasco:2021bmu}, taking its classical limit as in \cite{Georgoudis:2023lgf} and only evaluating its relevant permutations.

Using the fact that IR divergences come from exchanged gravitons with very small momentum \cite{Weinberg:1965nx}, one can show that the IR pole of $W_1^{\mu\nu}$ (which only appears in its imaginary part) factorizes as the tree-level amplitude times a simple function, and can be isolated as follows\footnote{We drop the explicit wavepacket average, since we are now dealing with classical objects.}
\begin{equation}\label{newIR}
	W_1
	=
	-\frac{i}{\epsilon} u_0(k) \mathcal A_0 + \mathcal B_1 + i \mathcal M_1
\end{equation}
where $u_0(k)=u_0^s(k)+u_0^c(k)$ \cite{Caron-Huot:2023vxl}
\begin{equation}\label{u0def}
	u_0^s(k) = -G( m_1 u_1 + m_2 u_2)\cdot k\, \frac{y\,(y^2-\frac{3}{2})}{(y^2-1)^{3/2}}
	\,,\qquad
	u_0^c(k) = -G( m_1 u_1 + m_2 u_2)\cdot k
\end{equation}
and 
$\mathcal M_1$ denotes the $\mathcal O(\epsilon^0)$ terms in the imaginary part of the KMOC kernel.
We have checked that the IR divergence of $\frac{i}{2}(s-s')$ equals $-\frac{i}{\epsilon}\,u_0^s(k)\,\mathcal A_0$, while $-\frac{i}{\epsilon}\,u_0^c(k)\,\mathcal A_0$ arises from $\frac{i}{2}(c_1+c_2)$ as emphasized in Refs.~\cite{Brandhuber:2023hhy,Herderschee:2023fxh,Elkhidir:2023dco,Georgoudis:2023lgf}, so that the full one-loop kernel reproduces the IR factorization displayed in \eqref{newIR}, \eqref{u0def}. When performing this check, we can drop $\mathcal O(\hbar^{-1})$ terms that are analytic in the momentum transfers $q_1$, $q_2$ appearing in the Fourier transform \eqref{FT5}, which would only contribute to short-range terms in impact-parameter space. We flag terms of this kind that arise from contact $2\to2$ sub-topologies in the cut with a parameter ``$a$'' in the ancillary files. 

Taking the asymptotic limit of large spatial distance $r$ from the scattering event at fixed retarded time $u$ and angular direction characterized by the null vector $n^\mu$ leads to the identification $k^\mu = \omega n^\mu$. IR divergences of the type \eqref{newIR} can be thus subtracted by redefining the retarded time $u \to  u - \frac{1}{\epsilon}\,u_0(n)$ \cite{Goldberger:2009qd,Porto:2012as},
and leave behind finite ambiguities of the form $i\,u_0(k)\,\mathcal A_0\,\log\mu_{\text{IR}}^2$.

Coming to the finite $\mathcal O(\epsilon^0)$ part, again denoting by $\mathcal M_1^s$ and $\mathcal M_1^c$ the parts arising from $\frac{i}{2}(s-s')$ and $\frac{i}{2}(c_1+c_2)$, we refer to \cite{Brandhuber:2023hhy,Herderschee:2023fxh,Georgoudis:2023lgf} for the explicit expression of $\mathcal M_1^c$ and discuss the new result for $\mathcal M_1^s$.
Letting $\omega_1=-u_1\cdot k$ and $\omega_2=-u_2\cdot k$, we find the following structure,
\begin{equation}\label{M1s}
	\begin{split}
\mathcal M_1^{s}
&=
 G\, m_1 \omega_1\, \frac{y\,(y^2-\frac{3}{2})}{(y^2-1)^{3/2}}\,\mathcal A_0\,
\log\frac{q_1^2 (y^2-1)}{\mu_\text{IR}^2}\\
&+
\frac{1}{\sqrt{y^2-1}}
\left(
M_\text{rat} 
+
M_1\,\log\frac{q_1^2}{\omega_1^2}
+
M_2\,\log\frac{q_1^2\omega_1^2}{(q_2^2)^2}
+
M_3\,\log\frac{q_1^2}{\omega_2^2}
\right)
\\
&
+
M_4\,\arccosh y
+
M_5\,
\frac{\log(y^2-1)}{\sqrt{y^2-1}} + (1\leftrightarrow2)\,,
	\end{split}
\end{equation}
where the first line is dictated by the IR divergence, as mentioned above, and the $M_i$ appearing in the second line are rational functions of $q_1^2$, $q_2^2$, $\omega_1$, $\omega_2$ and $y$ whose numerators are quadratic in $\varepsilon\cdot q_2$, $\varepsilon\cdot u_1$, $\varepsilon\cdot u_2$, with $\varepsilon\cdot k=0$, $\varepsilon\cdot \varepsilon=0$, $\varepsilon\cdot \varepsilon^\ast=1$. 

As a cross-check on the finite part of $\frac{i}{2}(s-s')$ displayed in \eqref{M1s}, we have verified that its leading contribution $\sim 1/\omega$ in the soft limit $\omega\to0$ correctly reconstructs the longitudinal contributions to the one-loop memory effect as shown in \cite{Aoude:2023dui}, and that a similar mechanism holds for the $\log\omega$ contribution dictated by \cite{Sahoo:2018lxl,Sahoo:2021ctw}. When taking $\omega\to0$, it is convenient to retain the analytic terms labeled by ``$a$'', since they render the $\omega$-scaling of the result manifestly $1/\omega$ by canceling other analytic pieces that arise at order $1/\omega^3$ and $1/\omega^2$, although of course only the non-analytic pieces contribute to the memory. We have also checked that the  denominators
\begin{equation}\label{}
	\mathcal{P}=-\omega_1^2+2 \omega_1 \omega_2 y-\omega_2^2\,,
	\qquad
	\mathcal{Q}=(q_1^2)^2\omega_1^2-2 q_1^2 q_2^2 \omega_1 \omega_2 y+(q_2^2)^2 \omega_2^2
\end{equation}
appearing in \eqref{M1s} 
do not give rise to any singularity upon suitably imposing vanishing of the Gram determinant $\operatorname{det}[\operatorname{Gram}(u_1,u_2,q_1,q_2,\varepsilon)]=0$ in four dimensions.
A further check of \eqref{M1s} to leading order in the post-Newtonian limit is discussed in \cite{Georgoudis:2023eke}.

\acknowledgments{
We are thankful to Lara Bohnenblust, Simon Caron-Huot, Stefano De Angelis, Radu Roiban, Rodolfo Russo and Fei Teng for useful discussions. We are especially grateful to Lara Bohnenblust for sharing preliminary results \cite{toapIta} and checking the agreement with ours.
A.~G.~is supported by a Royal Society funding, URF\textbackslash R\textbackslash221015. C.~H.~is supported
by UK Research and Innovation (UKRI) under the UK government’s Horizon Europe funding guarantee [grant EP/X037312/1]. 
I.~V.~H. is supported by the Knut and Alice Wallenberg Foundation under grants KAW
2018.0116 and KAW 2018.0162.
No new data were generated or analyzed during this study.
}

\providecommand{\href}[2]{#2}\begingroup\raggedright\endgroup


\begin{thebibliography}{10}
	
	\bibitem{Caron-Huot:2023vxl}
	S.~Caron-Huot, M.~Giroux, H.~S. Hannesdottir, and S.~Mizera, ``{What can be
		measured asymptotically?},'' \href{http://arxiv.org/abs/2308.02125}{{\tt
			arXiv:2308.02125 [hep-th]}}.
	
	\bibitem{Bjerrum-Bohr:2018xdl}
	N.~E.~J. Bjerrum-Bohr, P.~H. Damgaard, G.~Festuccia, L.~Plant{\'e}, and
	P.~Vanhove, ``{General Relativity from Scattering Amplitudes},''
	\href{http://dx.doi.org/10.1103/PhysRevLett.121.171601}{{\em Phys. Rev.
			Lett.} {\bf 121} (2018) no.~17, 171601},
	\href{http://arxiv.org/abs/1806.04920}{{\tt arXiv:1806.04920 [hep-th]}}.
	
	\bibitem{Cheung:2018wkq}
	C.~Cheung, I.~Z. Rothstein, and M.~P. Solon, ``{From Scattering Amplitudes to
		Classical Potentials in the Post-Minkowskian Expansion},''
	\href{http://dx.doi.org/10.1103/PhysRevLett.121.251101}{{\em Phys. Rev.
			Lett.} {\bf 121} (2018) no.~25, 251101},
	\href{http://arxiv.org/abs/1808.02489}{{\tt arXiv:1808.02489 [hep-th]}}.
	
	\bibitem{Kosower:2018adc}
	D.~A. Kosower, B.~Maybee, and D.~O'Connell, ``{Amplitudes, Observables, and
		Classical Scattering},''
	\href{http://dx.doi.org/10.1007/JHEP02(2019)137}{{\em JHEP} {\bf 02} (2019)
		137},
	\href{http://arxiv.org/abs/1811.10950}{{\tt arXiv:1811.10950 [hep-th]}}.
	
	\bibitem{Bern:2019nnu}
	Z.~Bern, C.~Cheung, R.~Roiban, C.-H. Shen, M.~P. Solon, and M.~Zeng,
	``{Scattering Amplitudes and the Conservative Hamiltonian for Binary Systems
		at Third Post-Minkowskian Order},''
	\href{http://dx.doi.org/10.1103/PhysRevLett.122.201603}{{\em Phys. Rev.
			Lett.} {\bf 122} (2019) no.~20, 201603},
	\href{http://arxiv.org/abs/1901.04424}{{\tt arXiv:1901.04424 [hep-th]}}.
	
	\bibitem{KoemansCollado:2019ggb}
	A.~Koemans~Collado, P.~Di~Vecchia, and R.~Russo, ``{Revisiting the second
		post-Minkowskian eikonal and the dynamics of binary black holes},''
	\href{http://dx.doi.org/10.1103/PhysRevD.100.066028}{{\em Phys. Rev. D} {\bf
			100} (2019) no.~6, 066028}, \href{http://arxiv.org/abs/1904.02667}{{\tt
			arXiv:1904.02667 [hep-th]}}.
	
	\bibitem{Bern:2019crd}
	Z.~Bern, C.~Cheung, R.~Roiban, C.-H. Shen, M.~P. Solon, and M.~Zeng, ``{Black
		Hole Binary Dynamics from the Double Copy and Effective Theory},''
	\href{http://dx.doi.org/10.1007/JHEP10(2019)206}{{\em JHEP} {\bf 10} (2019)
		206}, \href{http://arxiv.org/abs/1908.01493}{{\tt arXiv:1908.01493
			[hep-th]}}.
	
	\bibitem{Parra-Martinez:2020dzs}
	J.~Parra-Martinez, M.~S. Ruf, and M.~Zeng, ``{Extremal black hole scattering at
		$\mathcal{O}(G^3)$: graviton dominance, eikonal exponentiation, and
		differential equations},''
	\href{http://dx.doi.org/10.1007/JHEP11(2020)023}{{\em JHEP} {\bf 11} (2020)
		023}, \href{http://arxiv.org/abs/2005.04236}{{\tt arXiv:2005.04236
			[hep-th]}}.
	
	\bibitem{DiVecchia:2020ymx}
	P.~Di~Vecchia, C.~Heissenberg, R.~Russo, and G.~Veneziano, ``{Universality of
		ultra-relativistic gravitational scattering},''
	\href{http://dx.doi.org/10.1016/j.physletb.2020.135924}{{\em Phys. Lett. B}
		{\bf 811} (2020)  135924}, \href{http://arxiv.org/abs/2008.12743}{{\tt
			arXiv:2008.12743 [hep-th]}}.
	
	\bibitem{DiVecchia:2021ndb}
	P.~Di~Vecchia, C.~Heissenberg, R.~Russo, and G.~Veneziano, ``{Radiation
		Reaction from Soft Theorems},''
	\href{http://dx.doi.org/10.1016/j.physletb.2021.136379}{{\em Phys. Lett. B}
		{\bf 818} (2021)  136379}, \href{http://arxiv.org/abs/2101.05772}{{\tt
			arXiv:2101.05772 [hep-th]}}.
	
	\bibitem{Bern:2021dqo}
	Z.~Bern, J.~Parra-Martinez, R.~Roiban, M.~S. Ruf, C.-H. Shen, M.~P. Solon, and
	M.~Zeng, ``{Scattering Amplitudes and Conservative Binary Dynamics at ${\cal
			O}(G^4)$},'' \href{http://dx.doi.org/10.1103/PhysRevLett.126.171601}{{\em
			Phys. Rev. Lett.} {\bf 126} (2021) no.~17, 171601},
	\href{http://arxiv.org/abs/2101.07254}{{\tt arXiv:2101.07254 [hep-th]}}.
	
	\bibitem{Herrmann:2021lqe}
	E.~Herrmann, J.~Parra-Martinez, M.~S. Ruf, and M.~Zeng, ``{Gravitational
		Bremsstrahlung from Reverse Unitarity},''
	\href{http://dx.doi.org/10.1103/PhysRevLett.126.201602}{{\em Phys. Rev.
			Lett.} {\bf 126} (2021) no.~20, 201602},
	\href{http://arxiv.org/abs/2101.07255}{{\tt arXiv:2101.07255 [hep-th]}}.
	
	\bibitem{DiVecchia:2021bdo}
	P.~Di~Vecchia, C.~Heissenberg, R.~Russo, and G.~Veneziano, ``{The eikonal
		approach to gravitational scattering and radiation at $ \mathcal{O}
		$(G$^{3}$)},'' \href{http://dx.doi.org/10.1007/JHEP07(2021)169}{{\em JHEP}
		{\bf 07} (2021)  169}, \href{http://arxiv.org/abs/2104.03256}{{\tt
			arXiv:2104.03256 [hep-th]}}.
	
	\bibitem{Herrmann:2021tct}
	E.~Herrmann, J.~Parra-Martinez, M.~S. Ruf, and M.~Zeng, ``{Radiative classical
		gravitational observables at $ \mathcal{O} $(G$^{3}$) from scattering
		amplitudes},'' \href{http://dx.doi.org/10.1007/JHEP10(2021)148}{{\em JHEP}
		{\bf 10} (2021)  148}, \href{http://arxiv.org/abs/2104.03957}{{\tt
			arXiv:2104.03957 [hep-th]}}.
	
	\bibitem{Heissenberg:2021tzo}
	C.~Heissenberg, ``{Infrared divergences and the eikonal exponentiation},''
	\href{http://dx.doi.org/10.1103/PhysRevD.104.046016}{{\em Phys. Rev. D} {\bf
			104} (2021) no.~4, 046016}, \href{http://arxiv.org/abs/2105.04594}{{\tt
			arXiv:2105.04594 [hep-th]}}.
	
	\bibitem{Bjerrum-Bohr:2021vuf}
	N.~E.~J. Bjerrum-Bohr, P.~H. Damgaard, L.~Plant{\'e}, and P.~Vanhove,
	``{Classical gravity from loop amplitudes},''
	\href{http://dx.doi.org/10.1103/PhysRevD.104.026009}{{\em Phys. Rev. D} {\bf
			104} (2021) no.~2, 026009}, \href{http://arxiv.org/abs/2104.04510}{{\tt
			arXiv:2104.04510 [hep-th]}}.
	
	\bibitem{Bjerrum-Bohr:2021din}
	N.~E.~J. Bjerrum-Bohr, P.~H. Damgaard, L.~Plant{\'e}, and P.~Vanhove, ``{The
		amplitude for classical gravitational scattering at third Post-Minkowskian
		order},'' \href{http://dx.doi.org/10.1007/JHEP08(2021)172}{{\em JHEP} {\bf
			08} (2021)  172}, \href{http://arxiv.org/abs/2105.05218}{{\tt
			arXiv:2105.05218 [hep-th]}}.
	
	\bibitem{Cristofoli:2021vyo}
	A.~Cristofoli, R.~Gonzo, D.~A. Kosower, and D.~O'Connell, ``{Waveforms from
		amplitudes},'' \href{http://dx.doi.org/10.1103/PhysRevD.106.056007}{{\em
			Phys. Rev. D} {\bf 106} (2022) no.~5, 056007},
	\href{http://arxiv.org/abs/2107.10193}{{\tt arXiv:2107.10193 [hep-th]}}.
	
	\bibitem{Damgaard:2021ipf}
	P.~H. Damgaard, L.~Plant{\'e}, and P.~Vanhove, ``{On an exponential
		representation of the gravitational S-matrix},''
	\href{http://dx.doi.org/10.1007/JHEP11(2021)213}{{\em JHEP} {\bf 11} (2021)
		213}, \href{http://arxiv.org/abs/2107.12891}{{\tt arXiv:2107.12891
			[hep-th]}}.
	
	\bibitem{Brandhuber:2021eyq}
	A.~Brandhuber, G.~Chen, G.~Travaglini, and C.~Wen, ``{Classical gravitational
		scattering from a gauge-invariant double copy},''
	\href{http://dx.doi.org/10.1007/JHEP10(2021)118}{{\em JHEP} {\bf 10} (2021)
		118}, \href{http://arxiv.org/abs/2108.04216}{{\tt arXiv:2108.04216
			[hep-th]}}.
	
	\bibitem{Bjerrum-Bohr:2021wwt}
	N.~E.~J. Bjerrum-Bohr, L.~Plant{\'e}, and P.~Vanhove, ``{Post-Minkowskian
		radial action from soft limits and velocity cuts},''
	\href{http://dx.doi.org/10.1007/JHEP03(2022)071}{{\em JHEP} {\bf 03} (2022)
		071}, \href{http://arxiv.org/abs/2111.02976}{{\tt arXiv:2111.02976
			[hep-th]}}.
	
	\bibitem{Bern:2021yeh}
	Z.~Bern, J.~Parra-Martinez, R.~Roiban, M.~S. Ruf, C.-H. Shen, M.~P. Solon, and
	M.~Zeng, ``{Scattering Amplitudes, the Tail Effect, and Conservative Binary
		Dynamics at ${\cal O}(G^4)$},''
	\href{http://dx.doi.org/10.1103/PhysRevLett.128.161103}{{\em Phys. Rev.
			Lett.} {\bf 128} (2022) no.~16, 161103},
	\href{http://arxiv.org/abs/2112.10750}{{\tt arXiv:2112.10750 [hep-th]}}.
	
	\bibitem{Cristofoli:2021jas}
	A.~Cristofoli, R.~Gonzo, N.~Moynihan, D.~O'Connell, A.~Ross, M.~Sergola, and
	C.~D. White, ``{The Uncertainty Principle and Classical Amplitudes},''
	\href{http://arxiv.org/abs/2112.07556}{{\tt arXiv:2112.07556 [hep-th]}}.
	
	\bibitem{Manohar:2022dea}
	A.~V. Manohar, A.~K. Ridgway, and C.-H. Shen, ``{Radiated Angular Momentum and
		Dissipative Effects in Classical Scattering},''
	\href{http://dx.doi.org/10.1103/PhysRevLett.129.121601}{{\em Phys. Rev.
			Lett.} {\bf 129} (2022) no.~12, 121601},
	\href{http://arxiv.org/abs/2203.04283}{{\tt arXiv:2203.04283 [hep-th]}}.
	
	\bibitem{DiVecchia:2022owy}
	P.~Di~Vecchia, C.~Heissenberg, and R.~Russo, ``{Angular momentum of
		zero-frequency gravitons},''
	\href{http://dx.doi.org/10.1007/JHEP08(2022)172}{{\em JHEP} {\bf 08} (2022)
		172}, \href{http://arxiv.org/abs/2203.11915}{{\tt arXiv:2203.11915
			[hep-th]}}.
	
	\bibitem{DiVecchia:2022nna}
	P.~Di~Vecchia, C.~Heissenberg, R.~Russo, and G.~Veneziano, ``{The eikonal
		operator at arbitrary velocities I: the soft-radiation limit},''
	\href{http://dx.doi.org/10.1007/JHEP07(2022)039}{{\em JHEP} {\bf 07} (2022)
		039}, \href{http://arxiv.org/abs/2204.02378}{{\tt arXiv:2204.02378
			[hep-th]}}.
	
	\bibitem{DiVecchia:2022piu}
	P.~Di~Vecchia, C.~Heissenberg, R.~Russo, and G.~Veneziano, ``{Classical
		Gravitational Observables from the Eikonal Operator},''
	\href{http://arxiv.org/abs/2210.12118}{{\tt arXiv:2210.12118 [hep-th]}}.
	
	\bibitem{DiVecchia:2023frv}
	P.~Di~Vecchia, C.~Heissenberg, R.~Russo, and G.~Veneziano, ``{The gravitational
		eikonal: from particle, string and brane collisions to black-hole
		encounters},'' \href{http://arxiv.org/abs/2306.16488}{{\tt arXiv:2306.16488
			[hep-th]}}.
	
	\bibitem{Damgaard:2023vnx}
	P.~H. Damgaard, E.~R. Hansen, L.~Plant\'e, and P.~Vanhove, ``{The relation
		between KMOC and worldline formalisms for classical gravity},''
	\href{http://dx.doi.org/10.1007/JHEP09(2023)059}{{\em JHEP} {\bf 09} (2023)
		059}, \href{http://arxiv.org/abs/2306.11454}{{\tt arXiv:2306.11454
			[hep-th]}}.
	
	\bibitem{Damgaard:2023ttc}
	P.~H. Damgaard, E.~R. Hansen, L.~Plant\'e, and P.~Vanhove, ``{Classical
		observables from the exponential representation of the gravitational
		S-matrix},'' \href{http://dx.doi.org/10.1007/JHEP09(2023)183}{{\em JHEP} {\bf
			09} (2023)  183}, \href{http://arxiv.org/abs/2307.04746}{{\tt
			arXiv:2307.04746 [hep-th]}}.
	
	\bibitem{Goldberger:2016iau}
	W.~D. Goldberger and A.~K. Ridgway, ``{Radiation and the classical double copy
		for color charges},''
	\href{http://dx.doi.org/10.1103/PhysRevD.95.125010}{{\em Phys. Rev. D} {\bf
			95} (2017) no.~12, 125010}, \href{http://arxiv.org/abs/1611.03493}{{\tt
			arXiv:1611.03493 [hep-th]}}.
	
	\bibitem{Kalin:2020mvi}
	G.~K\"alin and R.~A. Porto, ``{Post-Minkowskian Effective Field Theory for
		Conservative Binary Dynamics},''
	\href{http://dx.doi.org/10.1007/JHEP11(2020)106}{{\em JHEP} {\bf 11} (2020)
		106}, \href{http://arxiv.org/abs/2006.01184}{{\tt arXiv:2006.01184
			[hep-th]}}.
	
	\bibitem{Kalin:2020fhe}
	G.~K\"alin, Z.~Liu, and R.~A. Porto, ``{Conservative Dynamics of Binary Systems
		to Third Post-Minkowskian Order from the Effective Field Theory Approach},''
	\href{http://dx.doi.org/10.1103/PhysRevLett.125.261103}{{\em Phys. Rev.
			Lett.} {\bf 125} (2020) no.~26, 261103},
	\href{http://arxiv.org/abs/2007.04977}{{\tt arXiv:2007.04977 [hep-th]}}.
	
	\bibitem{Mogull:2020sak}
	G.~Mogull, J.~Plefka, and J.~Steinhoff, ``{Classical black hole scattering from
		a worldline quantum field theory},''
	\href{http://dx.doi.org/10.1007/JHEP02(2021)048}{{\em JHEP} {\bf 02} (2021)
		048}, \href{http://arxiv.org/abs/2010.02865}{{\tt arXiv:2010.02865
			[hep-th]}}.
	
	\bibitem{Jakobsen:2021smu}
	G.~U. Jakobsen, G.~Mogull, J.~Plefka, and J.~Steinhoff, ``{Classical
		Gravitational Bremsstrahlung from a Worldline Quantum Field Theory},''
	\href{http://dx.doi.org/10.1103/PhysRevLett.126.201103}{{\em Phys. Rev.
			Lett.} {\bf 126} (2021) no.~20, 201103},
	\href{http://arxiv.org/abs/2101.12688}{{\tt arXiv:2101.12688 [gr-qc]}}.
	
	\bibitem{Mougiakakos:2021ckm}
	S.~Mougiakakos, M.~M. Riva, and F.~Vernizzi, ``{Gravitational Bremsstrahlung in
		the post-Minkowskian effective field theory},''
	\href{http://dx.doi.org/10.1103/PhysRevD.104.024041}{{\em Phys. Rev. D} {\bf
			104} (2021) no.~2, 024041}, \href{http://arxiv.org/abs/2102.08339}{{\tt
			arXiv:2102.08339 [gr-qc]}}.
	
	\bibitem{Dlapa:2021npj}
	C.~Dlapa, G.~K\"alin, Z.~Liu, and R.~A. Porto, ``{Dynamics of binary systems to
		fourth Post-Minkowskian order from the effective field theory approach},''
	\href{http://dx.doi.org/10.1016/j.physletb.2022.137203}{{\em Phys. Lett. B}
		{\bf 831} (2022)  137203}, \href{http://arxiv.org/abs/2106.08276}{{\tt
			arXiv:2106.08276 [hep-th]}}.
	
	\bibitem{Riva:2021vnj}
	M.~M. Riva and F.~Vernizzi, ``{Radiated momentum in the post-Minkowskian
		worldline approach via reverse unitarity},''
	\href{http://dx.doi.org/10.1007/JHEP11(2021)228}{{\em JHEP} {\bf 11} (2021)
		228}, \href{http://arxiv.org/abs/2110.10140}{{\tt arXiv:2110.10140
			[hep-th]}}.
	
	\bibitem{Dlapa:2021vgp}
	C.~Dlapa, G.~K\"alin, Z.~Liu, and R.~A. Porto, ``{Conservative Dynamics of
		Binary Systems at Fourth Post-Minkowskian Order in the Large-Eccentricity
		Expansion},'' \href{http://dx.doi.org/10.1103/PhysRevLett.128.161104}{{\em
			Phys. Rev. Lett.} {\bf 128} (2022) no.~16, 161104},
	\href{http://arxiv.org/abs/2112.11296}{{\tt arXiv:2112.11296 [hep-th]}}.
	
	\bibitem{Jakobsen:2022psy}
	G.~U. Jakobsen, G.~Mogull, J.~Plefka, and B.~Sauer, ``{All things retarded:
		radiation-reaction in worldline quantum field theory},''
	\href{http://dx.doi.org/10.1007/JHEP10(2022)128}{{\em JHEP} {\bf 10} (2022)
		128}, \href{http://arxiv.org/abs/2207.00569}{{\tt arXiv:2207.00569
			[hep-th]}}.
	
	\bibitem{Kalin:2022hph}
	G.~K\"alin, J.~Neef, and R.~A. Porto, ``{Radiation-Reaction in the Effective
		Field Theory Approach to Post-Minkowskian Dynamics},''
	\href{http://arxiv.org/abs/2207.00580}{{\tt arXiv:2207.00580 [hep-th]}}.
	
	\bibitem{Dlapa:2022lmu}
	C.~Dlapa, G.~K\"alin, Z.~Liu, J.~Neef, and R.~A. Porto, ``{Radiation Reaction
		and Gravitational Waves at Fourth Post-Minkowskian Order},''
	\href{http://arxiv.org/abs/2210.05541}{{\tt arXiv:2210.05541 [hep-th]}}.
	
	\bibitem{Dlapa:2023hsl}
	C.~Dlapa, G.~K\"alin, Z.~Liu, and R.~A. Porto, ``{Bootstrapping the
		relativistic two-body problem},''
	\href{http://dx.doi.org/10.1007/JHEP08(2023)109}{{\em JHEP} {\bf 08} (2023)
		109}, \href{http://arxiv.org/abs/2304.01275}{{\tt arXiv:2304.01275
			[hep-th]}}.
	
	\bibitem{Brandhuber:2023hhy}
	A.~Brandhuber, G.~R. Brown, G.~Chen, S.~De~Angelis, J.~Gowdy, and
	G.~Travaglini, ``{One-loop gravitational bremsstrahlung and waveforms from a
		heavy-mass effective field theory},''
	\href{http://dx.doi.org/10.1007/JHEP06(2023)048}{{\em JHEP} {\bf 06} (2023)
		048}, \href{http://arxiv.org/abs/2303.06111}{{\tt arXiv:2303.06111
			[hep-th]}}.
	
	\bibitem{Herderschee:2023fxh}
	A.~Herderschee, R.~Roiban, and F.~Teng, ``{The sub-leading scattering waveform
		from amplitudes},'' \href{http://dx.doi.org/10.1007/JHEP06(2023)004}{{\em
			JHEP} {\bf 06} (2023)  004}, \href{http://arxiv.org/abs/2303.06112}{{\tt
			arXiv:2303.06112 [hep-th]}}.
	
	\bibitem{Elkhidir:2023dco}
	A.~Elkhidir, D.~O'Connell, M.~Sergola, and I.~A. Vazquez-Holm, ``{Radiation and
		Reaction at One Loop},'' \href{http://arxiv.org/abs/2303.06211}{{\tt
			arXiv:2303.06211 [hep-th]}}.
	
	\bibitem{Georgoudis:2023lgf}
	A.~Georgoudis, C.~Heissenberg, and I.~Vazquez-Holm, ``{Inelastic exponentiation
		and classical gravitational scattering at one loop},''
	\href{http://dx.doi.org/10.1007/JHEP06(2023)126}{{\em JHEP} {\bf 06} (2023)
		126}, \href{http://arxiv.org/abs/2303.07006}{{\tt arXiv:2303.07006
			[hep-th]}}.
	
	\bibitem{Bini:2023fiz}
	D.~Bini, T.~Damour, and A.~Geralico, ``{Comparing One-loop Gravitational
		Bremsstrahlung Amplitudes to the Multipolar-Post-Minkowskian Waveform},''
	\href{http://arxiv.org/abs/2309.14925}{{\tt arXiv:2309.14925 [gr-qc]}}.
	
	\bibitem{Aoude:2023dui}
	R.~Aoude, K.~Haddad, C.~Heissenberg, and A.~Helset, ``{Leading-order
		gravitational radiation to all spin orders},''
	\href{http://arxiv.org/abs/2310.05832}{{\tt arXiv:2310.05832 [hep-th]}}.
	
	\bibitem{Georgoudis:2023eke}
	A.~Georgoudis, C.~Heissenberg, and R.~Russo, ``{An eikonal-inspired approach to
		the gravitational scattering waveform},''
	\href{http://arxiv.org/abs/2312.07452}{{\tt arXiv:2312.07452 [hep-th]}}.
	
	\bibitem{Carrasco:2021bmu}
	J.~J.~M. Carrasco and I.~A. Vazquez-Holm, ``{Extracting Einstein from the
		loop-level double-copy},''
	\href{http://dx.doi.org/10.1007/JHEP11(2021)088}{{\em JHEP} {\bf 11} (2021)
		088}, \href{http://arxiv.org/abs/2108.06798}{{\tt arXiv:2108.06798
			[hep-th]}}.
	
	\bibitem{Weinberg:1965nx}
	S.~Weinberg, ``{Infrared photons and gravitons},''
	\href{http://dx.doi.org/10.1103/PhysRev.140.B516}{{\em Phys. Rev.} {\bf 140}
		(1965)  B516--B524}.
	
	\bibitem{Goldberger:2009qd}
	W.~D. Goldberger and A.~Ross, ``{Gravitational radiative corrections from
		effective field theory},''
	\href{http://dx.doi.org/10.1103/PhysRevD.81.124015}{{\em Phys. Rev. D} {\bf
			81} (2010)  124015}, \href{http://arxiv.org/abs/0912.4254}{{\tt
			arXiv:0912.4254 [gr-qc]}}.
	
	\bibitem{Porto:2012as}
	R.~A. Porto, A.~Ross, and I.~Z. Rothstein, ``{Spin induced multipole moments
		for the gravitational wave amplitude from binary inspirals to 2.5
		Post-Newtonian order},''
	\href{http://dx.doi.org/10.1088/1475-7516/2012/09/028}{{\em JCAP} {\bf 09}
		(2012)  028}, \href{http://arxiv.org/abs/1203.2962}{{\tt arXiv:1203.2962
			[gr-qc]}}.
	
	\bibitem{Sahoo:2018lxl}
	B.~Sahoo and A.~Sen, ``{Classical and Quantum Results on Logarithmic Terms in
		the Soft Theorem in Four Dimensions},''
	\href{http://dx.doi.org/10.1007/JHEP02(2019)086}{{\em JHEP} {\bf 02} (2019)
		086},
	\href{http://arxiv.org/abs/1808.03288}{{\tt arXiv:1808.03288 [hep-th]}}.
	
	\bibitem{Sahoo:2021ctw}
	B.~Sahoo and A.~Sen, ``{Classical soft graviton theorem rewritten},''
	\href{http://dx.doi.org/10.1007/JHEP01(2022)077}{{\em JHEP} {\bf 01} (2022)
		077}, \href{http://arxiv.org/abs/2105.08739}{{\tt arXiv:2105.08739
			[hep-th]}}.
	
	\bibitem{toapIta}
	L.~Bohnenblust, H.~Ita, M.~Kraus, and J.~Schlenk.
	\newblock To appear simultaneously.
	
\end{thebibliography}
\end{document}